# Demonstration of nonpolar *m*-plane vertical GaN-on-GaN p-n power diodes grown on free-standing GaN substrates

Houqiang Fu,[1] Xiaodong Zhang,[2] Kai Fu,[1] Hanxiao Liu,[3] Shanthan R. Alugubelli,[3] Xuanqi Huang,[1] Hong Chen,[1] Izak Baranowski,[1] Tsung-Han Yang,[1] Ke Xu,[2,4] Fernando A. Ponce,[3] Baoshun Zhang,[2] and Yuji Zhao[1,*]

[1]*School of Electrical, Computer and Energy Engineering, Arizona State University, Tempe, AZ 85287, USA*

[2]*Key Laboratory of Nanodevices and Applications, Suzhou Institute of Nano-tech and Nano-bionics, CAS, Suzhou 215123, People's Republic of China*

[3]*Department of Physics, Arizona State University, Tempe, AZ 85287, USA*

[4]*Suzhou Nanowin Science and Technology Co., Ltd., Suzhou, 215123, China*

*Abstract*

This work demonstrates the first nonpolar vertical GaN-on-GaN p-n power diodes grown on *m*-plane free-standing substrates by metalorganic chemical vapor deposition (MOCVD). The SEM and HRXRD results showed the good crystal quality of the homoepitaxial nonpolar structure with low defect densities. The CL result confirmed the nonpolar p-GaN was of high quality with well activated Mg acceptors and considerably reduced deep-level (DL) states. At forward bias, the device showed good rectifying behaviors with a turn-on voltage of 4.0 V, an on-resistance of 2.3 mΩ·cm$^2$, a high on/off ratio of ~ $10^{10}$, and a low leakage current of ~ $10^{-7}$ A/cm$^2$. At reverse bias, the current leakage and breakdown were described by the trap-assisted space-charge-limited current (SCLC) conduction mechanism, where *I* was proportional to $V^{4.5}$. The critical electrical field was calculated to be 2.0 MV/cm without field plates or edge termination, which is the highest value reported on nonpolar power devices. The high performance *m*-plane p-n diodes can serve as key building blocks to further develop nonpolar GaN power electronics and polarization-engineering-related advanced power device structures for power conversion applications.

Gallium nitride (GaN) has attracted considerable attention for efficient power conversion applications, due to its large bandgap, high breakdown electric field and high Baliga's figure of merit.[1] Previously, due to the lack of native GaN substrates, GaN power devices were usually heteroepitaxially grown on lattice-mismatched foreign substrates with large defect densities and limited device performance.[2,3] Recently, advancements in GaN crystal growth have made free-standing bulk GaN substrates commercially available.[4] Homoepitaxial growth of GaN power devices on these substrates can enable significant reduction in defect densities and improvements in device performance.[5,6] Due to the availability of heavily doped GaN substrates, novel vertical power devices have been developed for high voltage and high power applications.[1] In contrast with lateral devices such as high electron mobility transistors (HEMTs)[7], in vertical devices, the currents flow vertically through the devices and the reverse voltages are held vertically. The advantages of the vertical structure over the lateral structure are multifold[1]: smaller chip area, larger current, less sensitivity to surface states, better scalability, smaller current dispersion which is a big issue for HEMTs[7], etc.

Currently, the majority of vertical GaN-on-GaN power diodes such as p-n diodes[5,6] and transistors such as current aperture vertical electron transistors (CAVET)[8] have been demonstrated on polar *c*-plane substrates [Fig. 1(a)]. On the other hand, growing devices on nonpolar crystal orientations [Fig. 1(b)] can offer some unique material and device properties. *m*-plane is drastically different from *c*-plane in atomic arrangements [Fig. 1(c)]: *m*-plane consists of both Ga and N (1:1), while *c*-plane is exclusively comprised of Ga.[9] Therefore, the polar GaN *c*-plane features a very strong spontaneous and piezoelectric polarization fields.[10] This led to various issues in optoelectronics, such as quantum confined Stark effect (QCSE)[11] and efficiency droop[4,12]. The polarization-free nonpolar *m*-plane was proposed to address these issues.[13]

For power electronics, these polarization fields on the polar plane induce high density of two-dimensional electron gas (2DEG) for lateral HEMTs, which are promising for high frequency applications.[1,7] However, the 2DEG also makes it very difficult to realize a normally-off device that is highly desired for the safe circuit operation.[14] The nonpolar planes are advantageous for realizing normally-off transistors with large threshold voltages.[14,15] Some novel device concepts have also been theorized on *m*-plane such as vertical polarization-based superjunctions (SJs) and transistors.[16,17] In addition, there is also a pressing demand for more advanced GaN power device structures[18], such as vertical junction field-effect transistors (VJFETs)[19], junction barrier Schottky (JBS) diodes and SJs. The successful realization of these devices necessitates reliable selectively doped p-n junctions and a proper understanding of nonpolar p-n junctions [Fig. 1(d)]. Despite these advantages and potential impacts, to date there have been very limited reports on nonpolar power electronics and nonpolar vertical devices such as p-n diodes have still not been demonstrated due to the lack of *m*-plane n$^+$-GaN substrates and material challenges in growing high quality nonpolar epitaxial structures[4]. In this work, we reported for the first time nonpolar vertical GaN-on-GaN p-n diodes grown on free-standing substrates, where comprehensive material and device characterizations were performed.

The nonpolar p-n diodes were homoepitaxially grown on *m*-plane n$^+$-GaN free-standing substrates by metalorganic chemical vapor deposition (MOCVD). Trimethylgallium (TMGa) and ammonia (NH$_3$) were the source materials for Ga and N, respectively. Bis(cyclopentadienyl)magnesium (Cp$_2$Mg) and silane (SiH$_4$) were used as the precursors for p-type Mg dopants and n-type Si dopants, respectively. Hydrogen (H$_2$) was used as the carrier gas to transport the reactants to the heated substrates in the MOCVD reactor. More general information about the GaN growth by MOCVD can be found in Ref. 20. The device structure [Fig. 2(a)] was comprised of a 1.5 µm n$^+$-GaN buffer layer, a 4-µm-thick n-GaN drift layer, an 1 µm p-GaN layer, and a 20 nm p$^+$-GaN contact layer. Figure 2(b) shows the scanning electron microscopy (SEM) image of the cross-section of the device. The crystal quality of the epilayers was characterized by high resolution X-ray diffraction (HRXRD) using PANalytical X'Pert Pro system. The X-ray source is the Cu Kα radiation with a wavelength of 0.154 nm. Figure 2(c) shows the (200) plane rocking curves (RCs) of the sample. The full width at half maximum (FWHM) of the RC was 39 arc sec. The dislocation density of the sample was estimated to be in the high $10^6$ cm$^{-2}$ range according to Ref. 21.

The p-GaN of the nonpolar p-n diodes was also examined in detail by cathodoluminescence (CL), since p-doping in WBG semiconductors has always been a challenging task and p-GaN plays a critical role in the electrical properties of GaN devices. The CL spectrum was obtained using a JEOL 6300 SEM operating in the raster scan mode with an accelerating voltage of 7 kV and a beam current of 100 pA at room temperature. CL images were recorded by setting the monochromator to a specific wavelength to study the spatial distribution of the CL intensity. In Fig. 3(a), several peaks were identified in the CL spectrum using the Gaussian fitting. The 3.40 peak was very close the bandgap of GaN and was due to the free exciton transition near the band edge or the near band-edge (NBE) transition.[22] The 3.25 peak was caused by emission of the donor-acceptor pairs (DAP) due to the the conduction band (CB) or shallow donor to Mg acceptor transition.[23] The Mg activation energy was estimated to be ~ 150 meV. The 2.95 eV peak was due to the deep donor (Dd) to Mg acceptor transition. This peak has been widely observed and studied in p-GaN.[24] The deep donors were located ~ 0.30 eV below the CB and related to nitrogen vacancies ($V_N$). The formation of $V_N$ is energetically favorable in p-GaN and most of the Mg acceptors are compensated by $V_N$ especially at high Mg concentrations.[24] Therefore, $N_A$ in p-GaN is the net hole concentration after taking the self-compensation effect into account. In addition, we observed a very weak broad peak due to the deep level (DL) transition (i.e., the CB to deep acceptor transition[25]), indicating p-GaN on the nonpolar substrates had very few deep acceptors. The SE image in Fig. 3(b) showed a straited morphology featuring an undulated stripe-like surface, which is commonly observed in nonpolar and semipolar GaN and attributed to the in-plane diffusion anisotropy during the growth.[4] Figure 3(c)-3(e) indicated that there was no noticeable spatial variation in the CL intensity, suggesting a rather uniform p-GaN. These results showed that high quality p-GaN was obtained on the nonpolar substrates.

The device fabrication of nonpolar p-n diodes was completed using traditional optical photolithography, dry etching and metal lift-off processes. First, the samples were cleaned in acetone and isopropyl alcohol solvents and dipped in hydrochloric acid before metal depositions. Then, the mesa isolation of the devices was formed by the chlorine based inductively coupled plasma (ICP) dry etch with an etching depth of ~ 1 µm. The circular p-type ohmic contacts with a diameter of 200 µm were formed by Pd/Ni/Au metal stacks that were deposited by electron beam evaporation and subsequently annealed by rapid thermal annealing (RTA) at 450 °C for 5 minutes. For the n-type ohmic contact, non-alloyed Ti/Al/Ni/Au stacks were formed at the backside of the substrates using electron beam evaporation. No passivation, field plate (FP) or edge termination were employed in the devices. The capacitance-voltage (*C–V*) and forward current-voltage (*I–V*) characteristics were measured by the Keithley 4200-SCS parameter analyzer, and the reverse *I-V* characteristics by the Keithley 2410 sourcemeter.

Figure 4(a) presents the forward *I–V* characteristics of the nonpolar p-n diodes in linear scale. By extrapolating the leading edge of the *I-V* curve to the *x*-axis, the turn-on voltage ($V_{on}$) of the devices was determined to be ~ 4.0 V. The on-current of the device is limited by the upper current limit of the measurement setup, which is 0.1 A, not intrinsically by the devices. The nonpolar p-n diodes reached the current limit of 0.1 A at 4.9 V. Light emission from the devices was observed at forward bias beyond the $V_{on}$. It stems from the radiative recombination of injected electron and holes in the device drift layer. This usually indicates the good material quality of the epitaxial structure with low defect densities[26] since defects can serve as nonradiative recombination centers and quell the light emission. A home-made electroluminescence (EL) setup using a CCD spectrometer from Thorlabs was used to analyze the emission spectrum. As shown in Fig. 4(c), three peaks were observed at 367 nm (3.38 eV), 381 nm (3.25 eV), and 573 nm (2.16 eV), respectively. The first peak was due to the NBE emission, and the second peak was due the DAP emission. The third peak is due to the DL transition, indicating the existence of deep acceptors in the drift layer. The origin of the deep acceptors has been attributed to gallium vacancies ($V_{Ga}$) or interstitial carbon ($C_I$).[27]

The ideality factor $n$ of the devices was also extracted as a function of voltage by

$$n = \frac{q}{2.3kT} \frac{1}{d\log(J)/dV} \quad (1)$$

where $q$ is electron charge, $k$ is the Boltzmann constant, $T$ is temperature and $J$ is the current density. At bias below 2.0 V, the current is below the setup detection limit and $n$ couldn't be extracted. The $n$ first deceased to a minimum at a certain voltage and then increased with increasing voltage. The minimum $n$ of the p-n diodes was 2.1 at 2.50 V. The relatively high $n$ is due to the high p-GaN ohmic resistance.[3] Before 2.50 V, the transition from the Shockley-Read-Hall (SRH) recombination current to diode diffusion current resulted in the decrease of the $n$. The increase of the $n$ after 2.50 V is caused by series resistance effects.[28] In Fig. 4(b), current density and on-resistance ($R_{on}$) were plotted as a function of voltage in semi-log scale. The off-current density of the devices was below $10^{-7}$ A/cm$^2$, limited by the apparatus lower current limit of 0.1 nA. The device showed high on/off ratio on the order of ~ $10^{10}$, which is among the highest values demonstrated in vertical GaN p-n diodes.[5,10,26,29] At the current of 0.1 A (i.e., 0.3 kA/cm$^2$), the device had a $R_{on}$ of 2.3 mΩ·cm$^2$.

Figure 5(a) shows the temperature-dependent forward *I-V* characteristics of the nonpolar p-n diodes from 25°C to 175°C. The $V_{on}$ and $R_{on}$ were extracted as a function of temperature in Fig. 5(b). The $V_{on}$ decreased from 4.0 V to 3.7 V and the $R_{on}$ increased from 2.3 mΩ·cm$^2$ to 4.5 mΩ·cm$^2$ from 25 °C to 175 °C. The decrease of $V_{on}$ with increasing temperature was attributed to the exponentially increasing diode diffusion current.[28] By linearly fitting the experimental data, $V_{on}$ was found to decrease at a rate of 2.2 mV/°C. The increase of $R_{on}$ with temperature was due to the reduced carrier mobility at high temperature caused by strong phonon scattering.[6,30]

Figure 6(a) presents the reverse I–V characteristics of the nonpolar p-n diodes in log-log scale. A sudden change in the slope of the log I-log V curves of the devices was observed at a hump voltage of $V_{TFL}$ [traps-filled-limit (TFL) voltage] with a soft breakdown behavior.[3,32] When the reverse bias was below $V_{TFL}$, some of the injected electrons started to fill the deep acceptor traps and the rest contributed to the reverse leakage current. At the reverse bias of $V_{TFL}$, the sudden increase in the slope of the logarithmic I-V curves indicated that the electron-filling process of the deep acceptor traps was completed.[32] Such reverse characteristics can be described by a space-charge-limited current (SCLC) conduction mechanism with traps[33,34]

$$I = Aq^{1-l}\mu N_C \left\{\frac{\varepsilon_0 \varepsilon_r l}{N_T(l+1)}\right\}^l \left\{\frac{2l+1}{l+1}\right\}^{l+1} \frac{V^{l+1}}{d^{2l+1}} \quad (2)$$

where $A$ is the device area, $\mu$ is the mobility, $N_C$ is the effective density of states, $N_T$ is the trap density, $d$ is the thickness and $l = E_{CH}/kT$ where $E_{CH}$ is the characteristic energy of the exponential tail states in the sub-bandgap region caused by traps. Therefore, in the SCLC transport, $I$ is proportional to $V^m$ where $m = l + 1$. A good linear fitting was obtained between log I and log V in the logarithmic I-V curves for the devices where the coefficient of determination $R^2$ was close to unity. $m$ was found to be 4.5 and therefore the $E_{CH}$ was 90.3 meV. Figure 7(b) shows the electric field profile of the device according to the one-dimensional Poisson's equation. The $V_{TFL}$ was comprised of electric potential contributed by the ionized acceptors after the compensation effects, and ionized donors and charged deep acceptors traps in the p-GaN and n-GaN, respectively. The electric field $E_p$, $E_n$, and $V_{TFL}$ are expressed as

$$E_p = \frac{qN_A d_p}{\varepsilon_0 \varepsilon_r} \quad (3)$$

$$E_n = E_p - \frac{q(N_D - N_T)d_n}{\varepsilon_0 \varepsilon_r} \quad (4)$$

$$V_{TFL} = \frac{1}{2}E_p d_p + \frac{1}{2}(E_p + E_n)d_n = \frac{qN_A d_p^2}{2\varepsilon_0 \varepsilon_r} + \frac{q[2N_A d_p - (N_D - N_T)d_n]d_n}{2\varepsilon_0 \varepsilon_r} \quad (5)$$

where $d_p$ is the thickness of p-GaN and $d_n$ is the thickness of n-GaN. The critical electric field $E_c$ is the larger value between $E_p$ and $E_n$. The $N_D$ was extracted from C–V and $1/C^2$–V characteristics to be $3.5 \times 10^{17}$ cm$^{-3}$. The nonpolar p-n diode had a $V_{TFL}$ of 106 V, and $N_T$ was estimated to be 3.0

× $10^{17}$ cm$^{-3}$ according to Eq. 5. The $E_c$ was calculated to be 2.0 MV/cm for the devices—the record-high ever reported on nonpolar power devices. It is also comparable to the previously demonstrated polar p-n diodes without FP or edge termination,[10,21] though it was still lower the best values on polar p-n diodes (~ 3.0 MV/cm).

In summary, nonpolar vertical GaN p-n diodes were demonstrated on free standing *m*-plane free-standing GaN substrates by MOCVD. The material characterization results indicated good crystal quality of the *m*-plane epilayers with low defect densities and high quality p-GaN with greatly suppressed deep-level transition, which enabled excellent electrical properties of the nonpolar p-n diodes. Due to the fundamentally distinct polarization properties, nonpolar p-n junctions can not only provide additional device design freedom for existing device structures such as HEMTs and CAVET, but also be incorporated into other advanced power electronics structures such as VJFETs, JBS and SJs. This work has shown that high performance nonpolar p-n diodes can be obtained and readily used to support various novel device opportunities.

**Figures**

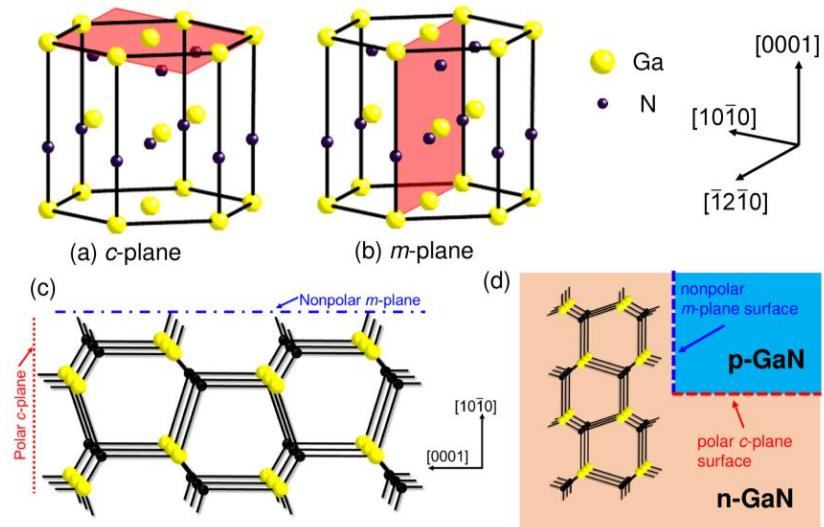

Figure 1 (a) Polar *c*-plane and (b) nonpolar *m*-plane in the GaN wurtzite crystal unit cell. (d) Atomic arrangements of *c*-plane and *m*-plane. (c) Schematics of selectively doped p-n junctions. During the fabrication of vertical JFETs JBS and SJs, the nonpolar surface will be exposed, and a nonpolar p-n junction will be formed.

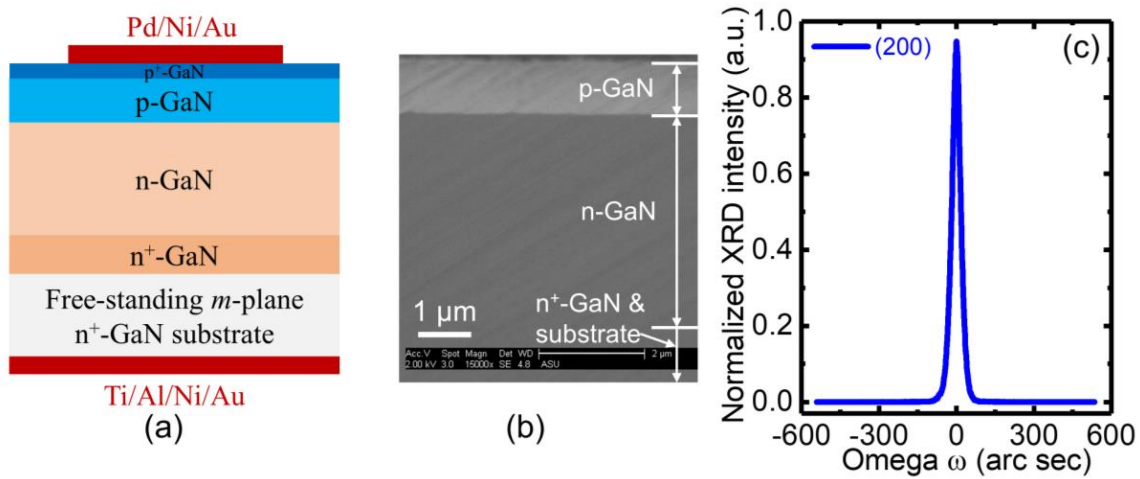

Figure 2 (a) Schematic view of the device structure of the nonpolar *m*-plane GaN p-n diodes. (b) The SEM image of the cross-section of the devices. (c) (200) plane rocking curve of the device epilayer by HRXRD.

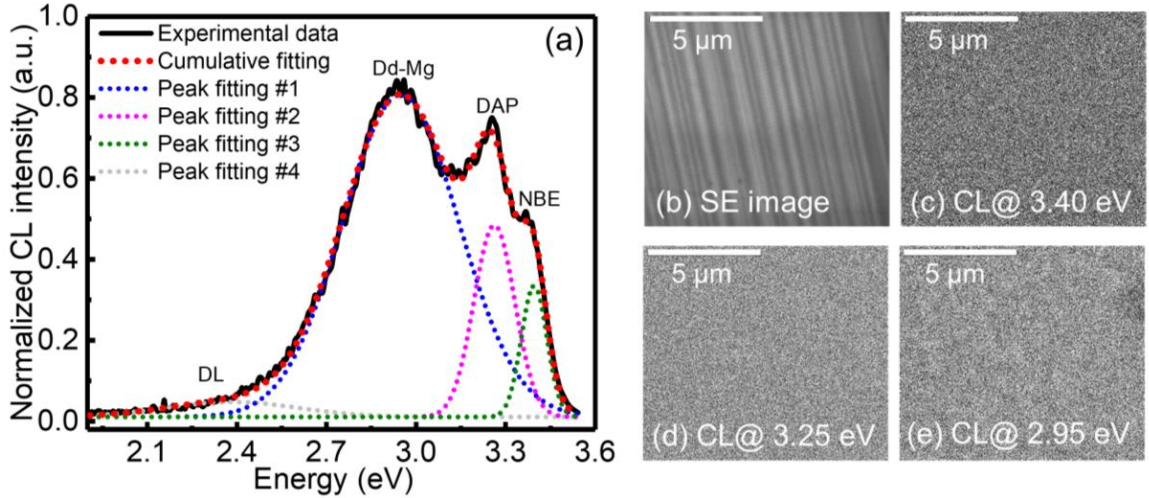

Figure 3 (a) CL spectrum of the p-GaN of the nonpolar p-n diodes. (c) SE image of the p-GaN. CL images of the p-GaN at (c) 3.40 eV, (d) 3.25 eV and (e) 2.95 eV.

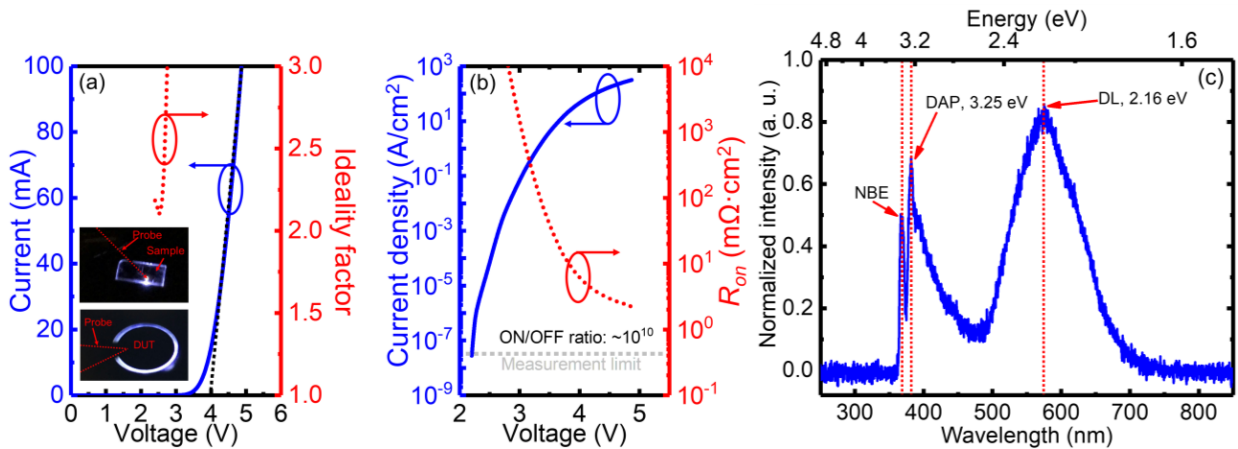

Figure 4 (a) Forward *I–V* and ideality factor as a function of voltage in linear scale. The inset shows the illuminated whole sample and a device under test (DUT) at bias beyond $V_{on}$. (b) Current density and $R_{on}$ vs. voltage in semi-log scale. (c) The EL spectrum of the luminescence from the device.

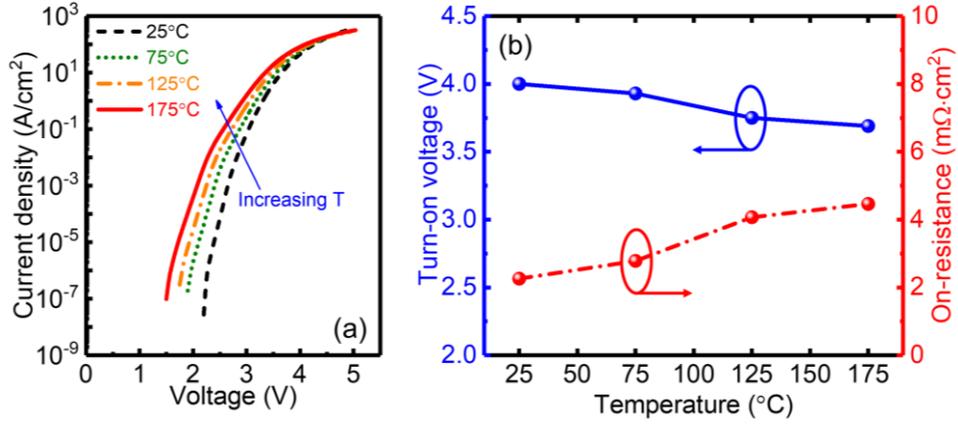

Figure 5 (a) Temperature dependent forward *I–V* characteristics from 25 °C to 175 °C. (b) The extracted $V_{on}$ and $R_{on}$ as a function of temperature.

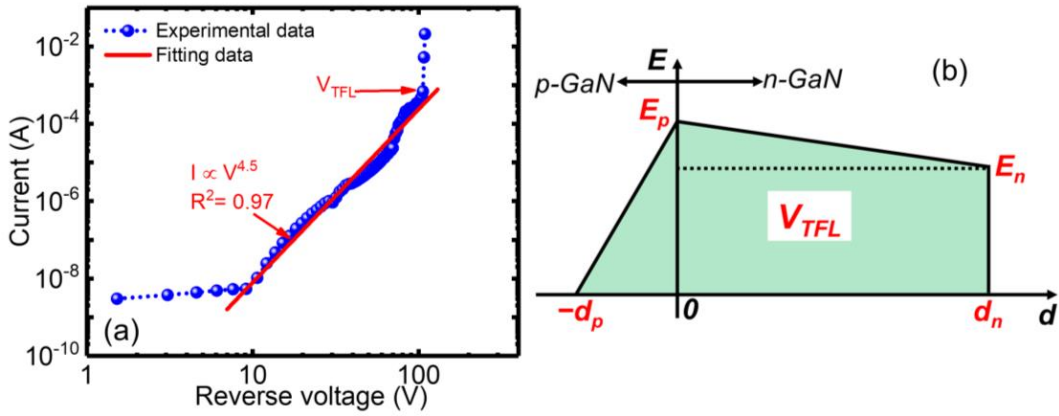

Figure 6 (a) Reverse *I–V* characteristics of the nonpolar p-n diodes in log-log scale. The data was fitted by the SCLC model. (b) Electric field profile along the growth direction of the devices.